# Improving Controllability of Complex Networks by Rewiring Links Regularly

Jiuqiang Xu, Jinfa Wang[*], Hai Zhao, Siyuan Jia

College of Information Science and Engineering,Northeastern University,Shenyang,110819,China
E-mail: mervin0502@163.com

**Abstract**: Network science have constantly been in the focus of research for the last decade, with considerable advances in the controllability of their structural. However, much less effort has been devoted to study that how to improve the controllability of complex networks. In this paper, a new algorithm is proposed to improve the controllability of complex networks by rewiring links regularly which transforms the network structure. Then it is demonstrated that our algorithm is very effective after numerical simulation experiment on typical network models (Erdös-Rényi and scale-free network). We find that our algorithm is mainly determined by the average degree and positive correlation of in-degree and out-degree of network and it has nothing to do with the network size. Furthermore, we analyze and discuss the correlation between controllability of complex networks and degree distribution index: power-law exponent and heterogeneity.

## 1 INTRODUNCTION

Controllability is an important property of control systems and plays a critical role in many control theory problems [1]. In the last decade, it has become the most challenging issue in modern network science. Many researchers have studied the structural controllability of complex network with linear dynamics and obtained some significant results[2]. Especially, after Liu et al.[2] develop a fundamental framework tool controlling the network based on the nodal dynamics, it is possible to analyze quantificationally the controllability of directed complex networks. Tamás et al.[4] introduce a new model controlling the dynamical edge to control the whole network. They find that positive correlation in-degrees and out-degrees improve the controllability of the complex networks. Yuan et al.[5] propose an exact controllability paradigm which introduce the maximum multiplicity to determine the minimum set of driver nodes to implement full control of arbitrary complex networks. Although they propose these tools or models for analyzing the controllability of complex networks, few people study that how to improve the controllability of complex networks[6].

Structural controllability[1] has been proposed in the Ref.2 as a framework tool to research the controllability method of directed complex networks[2]. The network with linear time-independent dynamics is depicted as follow in this framework,:

$$\dot{x} = \mathbf{A}x(t) + \mathbf{B}u(t) \qquad (1)$$

Where the vector $x(t) = (x_1(t), x_2(t), \cdots, x_N(t))^T$ stands for the state of nodes at time $t$, $\mathbf{A} \in R^{N \times N}$ denotes the transpose of the adjacency matrix of network, in which $a_{ij}$ represent the weight of one directed link from $j$ to $i$. The **B** is the $N \times m$ input matrix ($m \leq N$) which stands for how the $m$ input signals are connected to the network nodes, and the $u(t) = (u_1(t), u_2(t), \ldots, u_m(t))^T$ is a time-dependent vector of $m$ input signals. The classic Kalman rank condition[7] depicts that the linear system (equation (1)) is controllable no matter what state in finite time, iff the controllability matrix of $N \times Nm$

$$C = (\mathbf{B}, \mathbf{AB}, \mathbf{A^2B}, \cdots, \mathbf{A^{N-1}B}) \qquad (2)$$

has full rank, i.e. $rank(C) = N$. Noting that discussing the controllability of complex network in this paper is structural control. The nodes what are controlled directly by input signals $u(t)$ are called control nodes. The driver nodes is some of control nodes which are driven by different input signals. Controllability of network is characterized by the minimum set of driver nodes, which can offer full control over the network. If a single driver node can fully control the whole network, we say that the network has very perfect controllability.

We put forward a new algorithm to improve controllability of complex networks in this paper. The main idea of this algorithm is to delete the redundant link[2] and add one link between a maximal in-degree node and a maximal out-degree node if both are not connected. Then we introduce the degree correlation and degree distribution indexes to analyze our algorithm effective in this paper. The results of numerical simulation give us a new insight for better understanding the relationship between controllability and degree-correlation or degree-distribution in the complex networks.

## 2 REWIRING LINK ALGORITHM

Only if we control some nodes, we can control the whole network, so we hope to transform the network structure to reduce the number of driver nodes under ensuring the network size unchanged. The Ref.2 has classified the links into different types: critical link, redundant link and ordinary link. One link is critical, if removing it from the network increases the number of driver nodes to archive full control; one link is redundant, if its removal cannot affect the current set of driver nodes; the remaining links are ordinary. It is clear that removing the redundant links do not change the

This work was supported by the National Natural Science Foundation of China under grant 61101121.

driver nodes of current network, and adding a new link may improve the controllability of networks, but never reduce the controllability of network in the Liu model[2]. In the Ref.5, Hou et al. have proposed a rewiring link randomly algorithm: remove the redundant links and add the new links randomly. The controllability of network can be improved as long as the average degree of the network isn't less than 2.5 according to using Hou's algorithm. Hou's algorithm shows that it may have a good result only if it iterates itself many a time. However we attempt to find a new approach to improve the controllability of network here. In the Ref.4, the researchers indicate that positive correlation in-degrees and out-degrees improve the controllability of network. This is because the nodes of positive correlations network often represent complex decision processes which map a high-dimensional input space into a similarly high-dimensional output space. Strong positive correlations also yield networks with a higher number of short loops[9], which can be covered by closed walks that do not require driver nodes on their own.

Based on all above analysis, we put forward a new algorithm to improve controllability of complex networks—*Rewiring Links Regularly*. Rewiring links remains the network size and adding a new link between a node with maximal in-degree and a node with maximal out-degree if both are not connected increases the positive correlation. The *Rewiring Links Regularly* Algorithm is depicted as follows:

1. Apply the maximum matching theorem over the network to obtain the matched nodes and links[10].
2. Extract the redundant links by the filtering algorithm[11].
3. The algorithm stop, if the network has no redundant link.
4. Delete a redundant link.
5. Add one new link between maximal in-degree node and maximal out-degree one if both nodes are not connected.
6. Turn to step 3.

The complexity of maximum matching algorithm is at most $O(N^{1/2}L)$ for a directed network $G$ with $N$ nodes and $L$ links. The complexity of filtering algorithm is $O(N+L)$. The algorithm complexity of step 4 and 5 is $O(LN \lg N)$ in the worst case.

## 3 RESULTS and DISCUSSION

We studied the effect of our algorithm over two different type networks: Erdös-Rényi network[12] and Scale-free network[14]. In the Fig.1, we analyze the changes of density of driver nodes $n_D$ with different average degree $<k>$ of three networks: original network, produced by ER model or SF model with N=2000; Hou method, a new network produced by using Hou's method to original network; new method, a new network produced by using new method proposed in this paper to original network. It is clear that our new method has a better performance than Hou's method. But we also see different variation trend about the curve of "new method" in Fig.1 (a) (b). For Fig.1 (a), the number of driver nodes of ER network is approximately one when $<k>$ is greater than 4. The mainly reason is that the ER network is homogenous, and the dense and homogenous network is easier to control[2]. For Fig.1 (b), the density of driver nodes of original network slowly declines. This is because the sparse and heterogeneity network are very difficult to control[2]. While the number of driver nodes of new network produced by using our new method declines quickly and it is one when $<k>$ is greater than 5. It indicates that our method about improving the controllability of network is very effective. The more important is that it only needs to be run once.

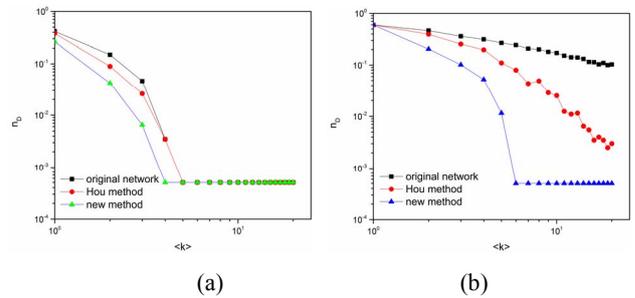

(a)        (b)

Fig.1 Controllability of the Erdös-Rényi and scale-free networks with different average degree values $<k>$. The density of driver nodes of network $n_D$ as a function of $<k>$. The power-law exponent of scale-free network γ is 4.

In our algorithm, we try to add a new link between maximal in-degree node and maximal out-degree node to increase the positive correlation between in-degrees and out-degrees of network. So as to observe the changes of network structure after rewiring links regularly, we introduce the definition of assortative coefficient of directed network[17] which is depicted by equation (3).

$$r(\alpha,\beta) = \frac{L^{-1}\sum_i [(j_i^\alpha - \overline{j}^\alpha)(k_i^\beta - \overline{k}^\beta)]}{\sigma^\alpha \sigma^\beta} \quad (3)$$

Where $\alpha,\beta \in \{in, out\}$ donte the degree type, $j_i^\alpha$ and $k_i^\beta$ is α-degree and β-degree of the source node and target node for edge $i$. $L$ is the number of edges in network, $\overline{j}^\alpha = L^{-1}\sum_i j_i^\alpha$ and $\sigma^\alpha = \sqrt{L^{-1}\sum_i (j_i^\alpha - \overline{j}^\alpha)^2}$ ; $\overline{k}^\beta$ and $\sigma^\beta$ are similarly defined. When $r(\alpha,\beta) > 0$, it indicates the tendency of nodes which high α-degree nodes tend to connect nodes with high β-degree. When $r(\alpha,\beta) < 0$, the nodes with high α-degree are more like to connect the nodes with small β-degree.

In the Fig.2, we find that the assortative coefficient of ER network is similar. The main reason is that every node of the ER network has similar in-degree and out-degree values, since the rand network model which has same probability to connect other node when it is created. The impact of network structure from rewiring links becomes weak. For the SF networks, we find the assortative coefficient of SF2 network is greater than SF1 network. It indicates that our algorithm has improved the positively correlated in-degrees and out-degrees. We also find that the assortative coefficient of new network (SF2) fluctuates between 0.1 and -0.1. The reason is that the SF network with different $<k>$ and $\gamma$ has different amounts of redundant links every time. But the assortative coefficient of all new networks (ER2 and SF2) produced by using our method always approach zero. This indicates that the homogonous network is easier to control than heterogeneity network which is disassortative or assortative. The result of Fig.2 confirms the Ref.4 conclusion again: positive correlation between in-degrees and out-degrees improve the controllability of network.

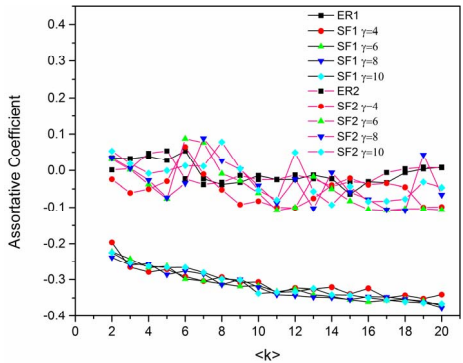

Fig. 2 Changes of assortative coefficient of networks between original network (ER1 and SF1) and new network (ER2 and SF2) produced by using our algorithm. Assortative coefficient as a function of $<k>$ for Erdös-Rényi (ER) and scale-free (SF) network with different power-law exponent $\gamma$. Here we compute the assortative coefficient of in- degree and out-degree of network $r(in, out)$ corresponding to the step 5 of our algorithm

We study the effect of network power-law exponent $\gamma$ and network size $N$ on improving the controllability of complex networks when we use our algorithm. In the Fig.3, it is obvious they have very similar results about the relationship between $<k>$ and $n_D$ with different network size N, i.e. the network size in our method can't affect network controllability. This is because that the network size doesn't change the degree distribution of network but only increases the computational complexity.

Fig.4 (a) shows that the relationship between $<k>$ and $n_D$ with different power-law exponent $\gamma$. When $<k>$ is between 2 and 4, the bigger the power-law exponent of network $\gamma$, the smaller the density of driver nodes $n_D$; when $<k>$ is greater than 5, the number of driver nodes of each network is one. It indicates that although the number of driver nodes of network is determined mainly by the network degree distribution[2], the redundant links will increase with increasing $<k>$. So we are going to rewiring more redundant links, while the impact of degree distribution with fixed $<k>$ become no longer important. It illustrates that our algorithm is very effective on transforming the network structure to improve the controllability of complex networks.

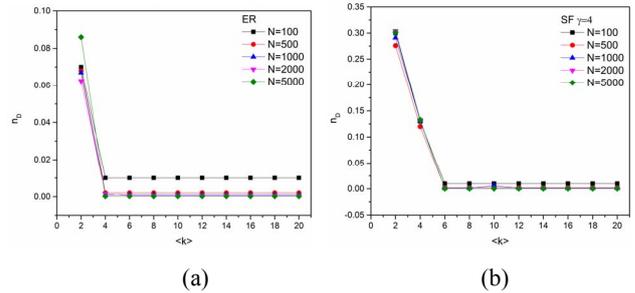

(a)          (b)

Fig.3 Controllability of the Erdös-Rényi and scale-free networks with different amount of node N. $n_D$ as a function of $<k>$.

Heterogeneity in the degree distribution is one of the most important complex network structure features[18,19]. Heterogeneity index provides a measure of the average degree inequality in a network, and larger index implies higher level of heterogeneity. The Ref.2 has shown that heterogeneity impact the controllability of complex networks: the less heterogeneous a network is, the better controllability it has[1].

According to analyzing the Fig.1 and Fig.4 (a), we find that although the controllability of network is improved by increasing the power-law exponent $\gamma$, when $<k>$ is larger than a certain value (it is 6 in the Fig.4), all networks can be controlled by only one node. So we wonder whether the increased heterogeneity must affect the controllability of network.

In order to knowing the impact of heterogeneity to controllability over the network, a calculate formula is introduced to calculate degree heterogeneity[21] $H$ here. It is depicted by equation (4).

$$H = \frac{\sum_{i}^{N}\sum_{j}^{N}|k_i - k_j|}{2N^2 <k>} \quad (4)$$

$H$ is one half of the relative average difference of all pairs of nodes. Where $N$ is the number of network nodes, $k_i$ denotes the degree of node $i$, $<k>$ denotes the average degree of network. Apparently, $0 \leqslant H < 1$, $H = 0$ for a completely homogeneous network, i.e. all nodes have the same degree value, and $H \to 1$ for a completely heterogeneous network. In the Fig.4 (b), we find that for the ER model, the original network and new network produced by our algorithm have very similar result. This is because that the ER network is homogonous network, as the Fig.1 (a) shows, it is completely controlled after using our algorithm when $<k>$ is equal to 4. For the SF networks with different power-law exponent $\gamma$, their heterogeneity index decrease slowly and smoothly. We find that the heterogeneity index new SF networks (SF2) have increased correspond to original networks'. Contacting to the Fig.1 results, it shows that even if our algorithm increases the heterogeneity index, the controllability of network not only doesn't decline but also been improved. It indicates that our algorithm is very effective. According to analyzing degree distribution index: power-law exponent and heterogeneity above all, it infers that there isn't a definitely negative correlation relationship between the controllability of complex network and degree distribution.

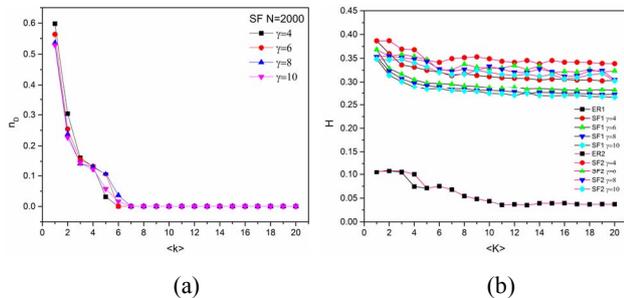

(a)　　　　　　　　(b)

Fig.4 The impact of network structure on the controllability of networks. (a) average degree, $<k>$, as a function of density of driver nodes of network $n_D$ with different values of $\gamma$. (b) average degree, $<k>$, as a function of heterogeneity $H$ for the Erdös-Rényi and scale-free network for variable $\gamma$.

## 4 CONCLUSION

In this paper, the *Rewiring Links Regularly Algorithm* is proposed to improve the controllability of complex networks. The main idea of algorithm proposed is to delete the redundant links and add new links between nodes with maximal in-degree and nodes with maximal out-degree if both are not connected. According to the algorithm, the purpose is not only to make more links act on the controllability of network but also to improve the positive correlation between in-degrees and out-degrees to enhance the controllability of complex network. We have found that this algorithm has a good performance in the Erdös-Rényi with different average degree and scale-free networks with different average degree or different power-law exponent. Furthermore, we study the correlation between network controllability and degree distribution index: power-law exponent and network heterogeneity, numerical stimulation shows that degree distribution of complex network do not impact the effectiveness of our algorithm in terms of improving the controllability of network.